\documentclass[aps,prl,twocolumn,superscriptaddress]{revtex4-2}
\usepackage{amsmath,amssymb,amsbsy,graphicx,xcolor,times,subfigure,marvosym,color,bm,multirow,epstopdf}
\usepackage[latin1]{inputenc}
\begin{document}

\title{Detecting half-quantum superconducting vortices by spin-qubit relaxometry}

\author{G\'abor B. Hal\'asz}
\thanks{This manuscript has been authored by UT-Battelle, LLC, under contract DE-AC05-00OR22725 with the US Department of Energy (DOE). The publisher acknowledges the US government license to provide public access under the DOE Public Access Plan (http://energy.gov/downloads/doe-public-access-plan).}
\affiliation{Materials Science and Technology Division, Oak Ridge National Laboratory, Oak Ridge, TN 37831, USA}

\author{Nirjhar Sarkar}
\affiliation{Materials Science and Technology Division, Oak Ridge National Laboratory, Oak Ridge, TN 37831, USA}

\author{Yueh-Chun Wu}
\affiliation{Materials Science and Technology Division, Oak Ridge National Laboratory, Oak Ridge, TN 37831, USA}

\author{Joshua T. Damron}
\affiliation{Chemical Sciences Division, Oak Ridge National Laboratory, Oak Ridge, TN 37831, USA}

\author{Chengyun Hua}
\affiliation{Materials Science and Technology Division, Oak Ridge National Laboratory, Oak Ridge, TN 37831, USA}

\author{Benjamin Lawrie}
\affiliation{Materials Science and Technology Division, Oak Ridge National Laboratory, Oak Ridge, TN 37831, USA}


\begin{abstract}

Half-quantum vortices in spin-triplet superconductors are predicted to harbor Majorana zero modes and may provide a viable avenue to topological quantum computation. Here, we introduce a novel approach for directly measuring the half-integer-quantized magnetic fluxes, $\Phi = h / (4e)$, carried by such half-quantum vortices via spin-qubit relaxometry. We consider a superconducting strip with a narrow pinch point at which vortices cross quasi-periodically below a spin qubit as a result of a bias current. We demonstrate that the relaxation rate of the spin qubit exhibits a pronounced peak if the vortex-crossing frequency matches the transition frequency of the spin qubit and conclude that the magnetic flux $\Phi$ of a single vortex can be obtained by dividing the corresponding voltage along the strip with the transition frequency. We discuss experimental constraints on implementing our proposed setup in spin-triplet candidate materials like UTe$_2$, UPt$_3$, and URhGe.

\end{abstract}


\maketitle


\emph{Introduction.}---Flux quantization is an essential hallmark of superconductivity that directly reflects the macroscopic coherence of the underlying quantum state. If a magnetic field is applied to a conventional type-II superconductor, the magnetic flux enters the superconducting material in quantized packets called vortices that each carry the exact same flux quantum $\Phi_0 = h / (2e)$. In exotic superconductors with spin-triplet pairing, however, fractional vortices carrying one half of the flux quantum, $\Phi_0 / 2$, have also been predicted~\cite{Ivanov-2001,Alicea-2012}. Remarkably, since these half-quantum vortices support Majorana zero modes~\cite{Kitaev-2001}, they provide a promising route to inherently fault-tolerant quantum computation~\cite{Kitaev-2003,Nayak-2008}. In turn, the spin-triplet superconducting parent state is potentially realized in materials like UTe$_2$~\cite{Ran-2019,Aoki-2022}, UPt$_3$~\cite{Fisher-1989,Tou-1998}, and URhGe~\cite{Aoki-2001,Levy-2005}. Nevertheless, the identification of spin-triplet superconductivity has proven to be a formidable task because standard experimental probes do not provide conclusive evidence~\cite{Mackenzie-2017}.

Quantum sensing with spin qubits has recently emerged as a powerful approach for locally probing both static and fluctuating magnetic fields in condensed-matter systems. Stereotypical spin qubits include nitrogen-vacancy (NV) centers in diamond~\cite{Rondin-2014,Casola-2018} and analogous boron vacancies in hexagonal boron nitride (hBN)~\cite{Gottscholl-2021,Liu-2022}. From the perspective of quantum sensing, these spin qubits leverage the quantum coherence and environmental sensitivity of isolated two-level systems to measure local magnetic fields with nanoscale spatial resolution~\cite{Degen-2017}. In the specific case of spin-qubit relaxometry, the longitudinal relaxation rate $1/T_1$ is sensitive to transverse magnetic-field fluctuations at the transition frequency of the spin qubit, thereby providing access to the magnetic-noise spectral density that may reflect microscopic dynamical correlations in a nearby condensed-matter system~\cite{Agarwal-2017,Rodriguez-Nieva-2018,Chatterjee-2019,Chatterjee-2022,Machado-2023,Potts-2025}. Experimentally, this new approach to quantum noise spectroscopy has been employed to probe spin dynamics in antiferromagnets~\cite{Finco-2021,Kumar-2024,Melendez-2025}, critical spin fluctuations near magnetic phase transitions~\cite{Wu-2025,Li-2025}, electrical noise in metals~\cite{Ariyaratne-2018} and superconductors~\cite{Monge-2023}, and even magnetic fluctuations induced by superconducting vortex dynamics~\cite{Jayaram-2025,Liu-2025}.

\begin{figure}[b]
\includegraphics[width=0.82\columnwidth]{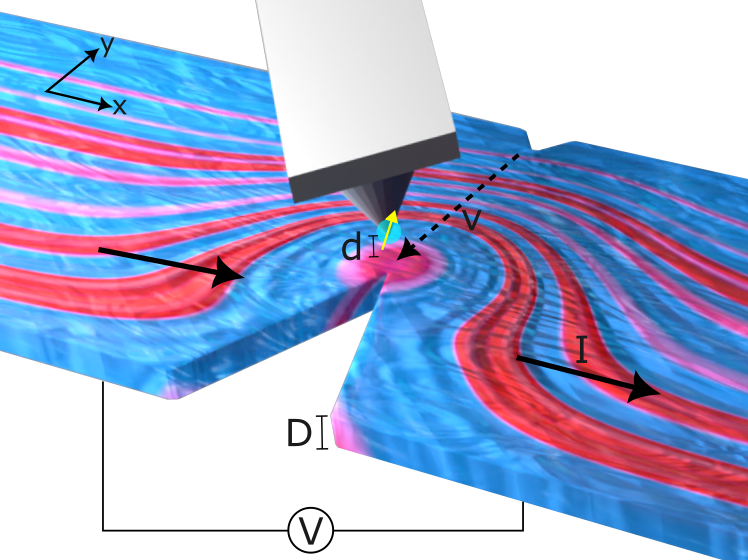}
\caption{Proposed setup for determining the magnetic flux $\Phi$ carried by a single superconducting vortex. The application of a sufficiently large bias current $I$ leads to quasi-periodic vortex crossing across the pinch point and, hence, a finite voltage $V$ along the superconducting strip of thickness $D$. Magnetic-field modulations induced by subsequent vortex-crossing events lead to a pronounced peak in the relaxation rate $1/T_1$ of the spin qubit at height $d$ above the strip when the vortex-crossing frequency $V / \Phi$ is commensurate with the transition frequency $f_0$ of the spin qubit. Each vortex nucleated at the smaller constriction on the top edge is focused toward the larger constriction at the bottom edge as a result of current crowding (red curves) and passes below the spin qubit with the same speed $v$.} \label{fig-1}
\end{figure}

In this Letter, we introduce and analyze a novel setup for detecting half-quantum vortices in spin-triplet superconductors via spin-qubit relaxometry. The proposed setup in Fig.~\ref{fig-1} relies on a narrow pinch point in a superconducting strip at which vortices cross quasi-periodically as a result of a finite bias current. We demonstrate that the relaxation rate $1/T_1$ of a spin qubit positioned directly above the pinch point shows a distinctive peak if the vortex-crossing frequency matches the transition frequency of the spin qubit. Moreover, the induced voltage along the strip is directly proportional to the vortex-crossing frequency, with the proportionality constant being the magnetic flux carried by a single vortex~\cite{Halperin-2010}. Hence, the single-vortex magnetic flux can be obtained as the ratio of the voltage corresponding to the above peak in $1/T_1$ and the transition frequency. Finding this magnetic flux to be $\Phi_0 / 2$ rather than $\Phi_0$ reveals that the vortices crossing the strip are half-quantum vortices with Majorana zero modes attached and that the underlying superconductivity is of spin-triplet nature.

\begin{figure*}[t]
\includegraphics[width=1.9\columnwidth]{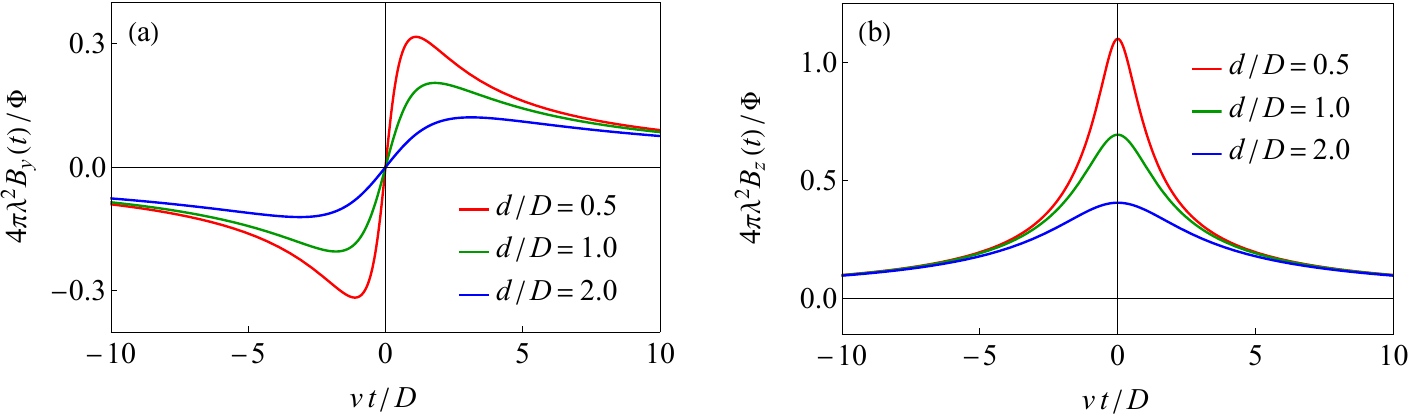}
\caption{Nonzero components (a) $B_y (t)$ and (b) $B_z (t)$ of the time-dependent magnetic field at the spin qubit during a single vortex-crossing event for different ratios between the distance $d$ of the spin qubit and the thickness $D$ of the superconductor.} \label{fig-2}
\end{figure*}

\emph{General idea.}---We consider a thin-film type-II superconducting strip with a small constriction on one edge, a much larger constriction on the opposite edge, and a spin qubit above the constriction region (see Fig.~\ref{fig-1}). In the presence of a small out-of-plane magnetic field, superconducting vortices start crossing the strip between the two constrictions once the bias current $I$ reaches a critical value. As the bias current increases further, the vortex-crossing frequency $f$ grows monotonically with the current $I$ and directly translates into a finite voltage $V = \Phi f$ along the superconducting strip~\cite{Halperin-2010}. Importantly, the magnetic flux $\Phi$ carried by a single vortex is $\Phi_0$ for standard superconducting vortices but $\Phi_0 / 2$ for half-quantum vortices hosted by spin-triplet superconductors.

If the bias current $I$ is applied in the appropriate direction, the vortices are nucleated at the small constriction~\cite{Bezuglyj-2022} and then move straight toward the large constriction that attracts them because of current crowding (see Fig.~\ref{fig-1}). Depending on the strip width, a new vortex is nucleated at the small constriction every time the previous vortex gets sufficiently far or is annihilated at the large constriction. In either case, we expect subsequent vortices to be separated by an approximately constant time difference $T = 1/f$. Regardless of the precise probability distribution, if the variation $\tau$ in this time difference is much smaller than the mean $T$, the quasi-periodic vortex crossing induces a time-dependent magnetic field at the spin qubit that has an approximate periodicity $T$. The dominant Fourier components of this quasi-periodic magnetic field then correspond to frequencies $n f$ (where $n$ is an integer), with $f$ being the fundamental frequency and $n f$ ($n \geq 2$) being higher harmonics. Hence, the relaxation rate $1 / T_1$ of the spin qubit is significantly increased whenever the vortex-crossing frequency $f$ is commensurate with the transition frequency $f_0$ of the spin qubit such that $f_0 = n f$. In other words, as the voltage $V$ is increased via the bias current $I$, the relaxation rate $1 / T_1$ exhibits distinctive peaks for a set of voltages, $V_n = \Phi f_0 / n$, and the magnetic flux $\Phi$ of a single vortex can be readily extracted as the ratio of the largest peak voltage $V_1$ and the transition frequency $f_0$.

\begin{figure*}[t]
\includegraphics[width=1.75\columnwidth]{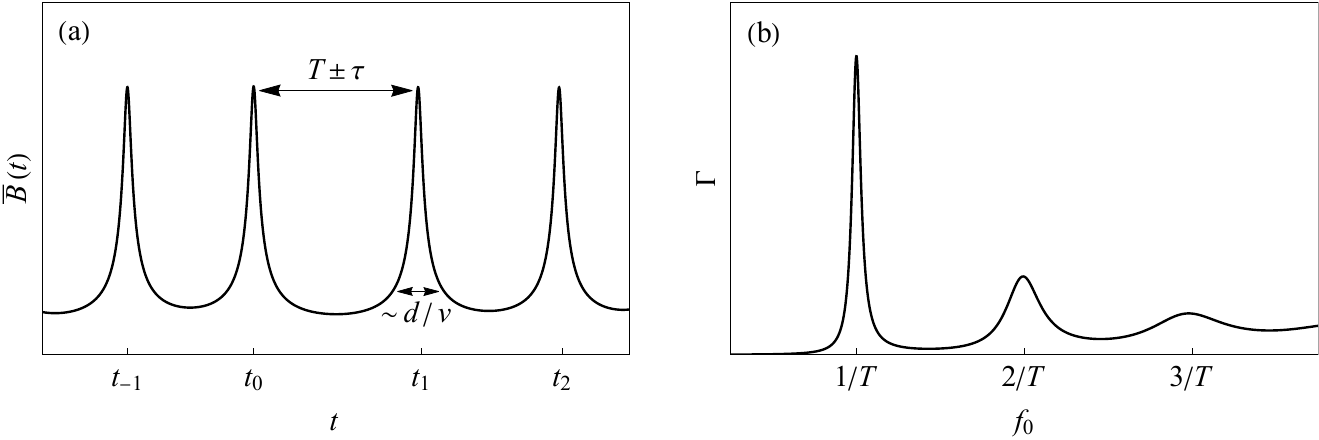}
\caption{(a) Schematic form of the time-dependent magnetic field $\bar{B} (t)$ at the spin qubit during a series of vortex-crossing events at $t = t_m$ separated by an approximately constant time difference $T \pm \tau$ with $\tau \ll T$. (b) Schematic form of the resulting relaxation rate $\Gamma \equiv 1/T_1$ against the transition frequency $f_0$ with a sequence of well-defined peaks at integer multiples of $1/T$.} \label{fig-3}
\end{figure*}

\emph{Magnetic field.}---We first obtain the time-dependent magnetic field at the spin qubit due to a single vortex crossing the superconducting strip between the two constrictions. We assume that the spin qubit is at distance $d$ above the top surface of the superconductor of thickness $D$ and that the path of the vortex goes directly through the vertical projection of the spin qubit (see Fig.~\ref{fig-1}). If the vortex moves in the $-y$ direction with constant speed $v$, the instantaneous Cartesian coordinates of the spin qubit with respect to the top of the vortex line are then $\vec{r}_0 = \{0, y_0, d\}$, where $y_0 = vt$ is the horizontal displacement as a function of time $t$. Also, the supercurrent density at a general position $\vec{r} = \{\rho \cos \varphi, \rho \sin \varphi, z\}$ inside the superconductor at $-D < z < 0$ is given by
\begin{equation}
\vec{j} (\vec{r}\,) = \frac{\Phi} {2\pi \mu_0 \lambda^2 \rho} \left\{ -\sin \varphi, \cos \varphi, 0 \right\}, \label{eq-j}
\end{equation}
where $\lambda \gg D$ is the London penetration depth, $\rho$ is the distance from the vortex line, and $\varphi$ is a cylindrical angle. Note that, while this expression is only valid for $\xi \ll r \ll \Lambda$,  where $\xi$ is the coherence length and $\Lambda = 2\lambda^2 / D$ is the Pearl length, the regions outside its applicability range have negligible contributions to the magnetic field as long as $\xi \ll d \ll \Lambda$. Using the Biot-Savart law, the magnetic field $\vec{B} = \{ B_x, B_y, B_z \}$ at the spin qubit then becomes
\begin{widetext}
\begin{equation}
\vec{B} = \frac{\mu_0} {4\pi} \int d^3 \vec{r} \,\, \frac{\vec{j} (\vec{r}\,) \times (\vec{r}_0 - \vec{r}\,)} {|\vec{r}_0 - \vec{r}\,|^3} = \frac{\Phi} {8\pi^2 \lambda^2} \int_{-D}^0 dz \int_0^{\infty} d\rho \, \rho \int_0^{2\pi} d\varphi \, \frac{\{(d-z) \cos \varphi, (d-z) \sin \varphi, \rho - y_0 \sin \varphi\}} {\rho \, [(d-z)^2 + \rho^2 + y_0^2 - 2 \rho y_0 \sin \varphi]^{3/2}}. \label{eq-B-1}
\end{equation}
\end{widetext}
The integrals in Eq.~(\ref{eq-B-1}) can be evaluated analytically, and the time-dependent field components with $y_0 = vt$ take the forms
\begin{equation}
B_{x,y,z} (t) = \frac{\Phi} {4\pi \lambda^2} \, K_{x,y,z} \left( \frac{d}{D}, \frac{vt}{D} \right), \label{eq-B-2}
\end{equation}
where $K_x (\gamma, \theta) = 0$ by symmetry, and the remaining dimensionless functions are given by
\begin{align}
K_y (\gamma, \theta) &= \frac{1 + \sqrt{\gamma^2 + \theta^2} - \sqrt{(1 + \gamma)^2 + \theta^2}} {\theta}, \nonumber \\
K_z (\gamma, \theta) &= \ln \left[ \frac{1 + \gamma + \sqrt{(1 + \gamma)^2 + \theta^2}} {\gamma + \sqrt{\gamma^2 + \theta^2}} \right]. \label{eq-K}
\end{align}
Figure~\ref{fig-2} shows the nonzero field components $B_{y,z} (t)$ against the dimensionless time $\theta = vt / D$ for various ratios $\gamma = d/D$.

\emph{Relaxation rate.}---In the next step, we determine the relaxation rate $\Gamma \equiv 1/T_1$ of the spin qubit due to quasi-periodic vortex crossing between the constrictions. If the intrinsic field of the spin qubit points in the $\vec{u}$ direction, its relaxation rate is sensitive to fluctuating magnetic fields in the two perpendicular directions $\vec{p}$ and $\vec{q}$. From Fermi's golden rule, the relaxation rate is then given by $\Gamma = \Gamma_p + \Gamma_q$ with~\cite{Chatterjee-2019}
\begin{equation}
\Gamma_p = \frac{\mu_B^2} {\hbar^2} \int_{-\infty}^{\infty} dt \, \langle \bar{B}_p (0) \bar{B}_p (t) \rangle \, e^{2\pi i f_0 t}, \label{eq-Gamma-1}
\end{equation}
where $f_0$ is the transition frequency of the spin qubit, while $\bar{B}_p (t)$ is the $p$ component of the magnetic field at the spin qubit induced by the quasi-periodic vortex motion. This time-dependent magnetic field, schematically sketched in Fig.~\ref{fig-3}(a), contains a sequence of identical features (e.g., peaks) that are centered at different times $t_m$ and correspond to crossings of individual vortices labeled by $m$. Neighboring features $m$ and $m+1$ are separated by an approximately constant time difference of mean $T$ and variation $\tau \ll T$, with the time $t_0$ of the $m = 0$ feature satisfying $-T/2 < t_0 < T/2$. We assume for simplicity that the time difference follows a Gaussian distribution but emphasize that our main results are not affected by this choice. Formally, the field component resulting from quasi-periodic vortex crossing can then be written as
\begin{align}
\bar{B}_p (t) = \sum_{m=-\infty}^{\infty} B_p (t - t_m), \quad \,\, t_m = t_0 + nT + \tau_m, \label{eq-B-3} \\
\nonumber
\end{align}
where $B_p (t)$ is the same field component due to the crossing of a single vortex [see Eq.~(\ref{eq-B-2}) and Fig.~\ref{fig-2}], while $t_0$ is a uniform random variable chosen with constant probability from the interval $-T/2 < t_0 < T/2$, and $\tau_m$ is a Gaussian random variable satisfying $\langle \tau_m \rangle = 0$ and $\langle \tau_m^2 \rangle = |m| \tau^2$. Note that the variation in $\tau_m$ scales with $\sqrt{|m|}$ because the variations of the $|m|$ time differences between $t_0$ and $t_m$ add in quadrature.

\begin{figure*}[t]
\includegraphics[width=2.0\columnwidth]{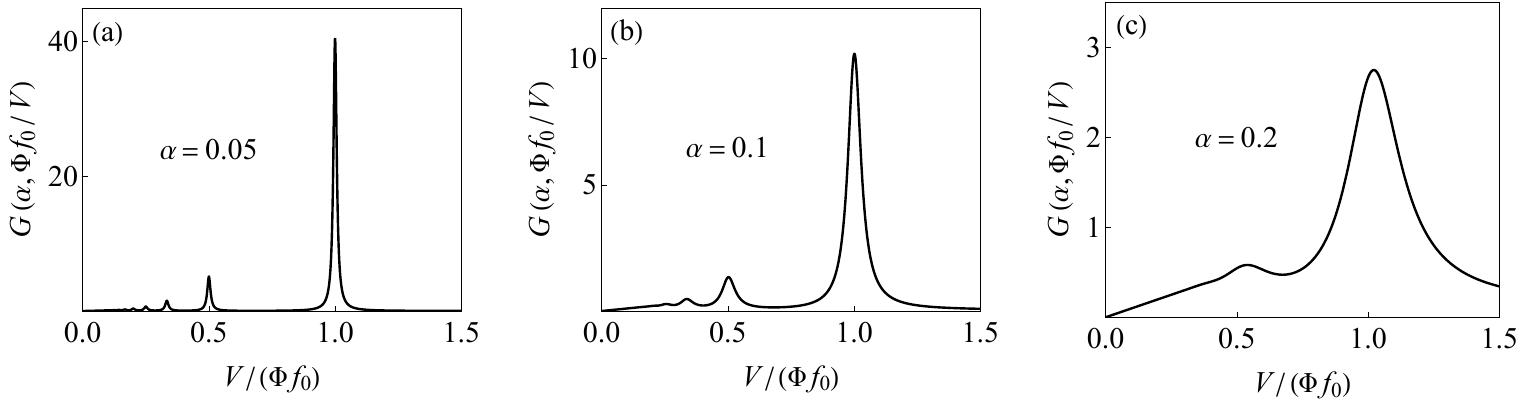}
\caption{Dimensionless function $G (\alpha, \Phi f_0 / V)$ against the voltage $V$ for different ratios $\alpha = \tau / T$ between the variation $\tau$ and the mean $T$ of the approximately constant time difference separating subsequent vortex-crossing events.} \label{fig-4}
\end{figure*}

Substituting Eq.~(\ref{eq-B-3}) into Eq.~(\ref{eq-Gamma-1}), the partial relaxation rate $\Gamma_p$ readily takes the form
\begin{widetext}
\begin{equation}
\Gamma_p = \frac{\mu_B^2} {\hbar^2 T} \int_{-\frac{T}{2}}^{\frac{T}{2}} dt_0 \int_{-\infty}^{\infty} dt \sum_{k=-\infty}^{\infty} \sum_{m=-\infty}^{\infty} \langle B_p (-t_0 - kT - \tau_k) B_p (t - t_0 - mT - \tau_m) \rangle \, e^{2\pi i f_0 t}, \label{eq-Gamma-2}
\end{equation}
\end{widetext}
where the averaging over the uniform random variable $t_0$ is explicitly included, and $\langle \ldots \rangle$ only denotes averaging over the Gaussian random variables $\tau_m$. If we assume $T \gg d/v$, the width of each identical feature in Fig.~\ref{fig-3}(a) is much smaller than the separation between neighboring features, and the sum over $k$ in Eq.~(\ref{eq-Gamma-2}) can be approximated with the $k = 0$ term alone. Introducing the new variable $t' = t - t_0 - mT - \tau_m$, the partial relaxation rate in Eq.~(\ref{eq-Gamma-2}) then becomes
\begin{widetext}
\begin{equation}
\Gamma_p \approx \frac{\mu_B^2} {\hbar^2 T} \sum_{m=-\infty}^{\infty} \left\langle e^{2\pi i f_0 (mT + \tau_m)} \right\rangle \int_{-\frac{T}{2}}^{\frac{T}{2}} dt_0 \, B_p (-t_0) \, e^{2\pi i f_0 t_0} \int_{-\infty}^{\infty} dt' B_p (t') \, e^{2\pi i f_0 t'}. \label{eq-Gamma-3}
\end{equation}
\end{widetext}
Using the standard identity $\langle e^{i x} \rangle = e^{-\frac{1}{2} \langle x^2 \rangle}$ for the Gaussian random variable $x = 2\pi f_0 \tau_m$, and taking the limit of $T \to \infty$ in the integration range (which is allowed because $T \gg d/v$), the partial relaxation rate can finally be written as
\begin{equation}
\Gamma_p \approx \frac{\mu_B^2 \Phi^2 D^2 f_0} {\hbar^2 \lambda^4 v^2} \, F_p \left( \frac{d}{D}, \frac{D f_0} {v} \right) G \bigg( \frac{\tau} {T}, f_0 T \bigg) \label{eq-Gamma-4}
\end{equation}
in terms of the dimensionless functions
\begin{align}
& F_p (\gamma, \phi) = \frac{1} {16\pi^2} \left| \int_{-\infty}^{\infty} d\theta \, K_p (\gamma, \theta) \, e^{2\pi i \phi \theta} \right|^2, \nonumber \\
& G (\alpha, \beta) = \frac{1} {\beta} \sum_{m=-\infty}^{\infty} e^{2\pi i m \beta - 2\pi^2 |m| \alpha^2 \beta^2} \label{eq-FG} \\
& \qquad \quad \,\, = \frac{\sinh (2\pi^2 \alpha^2 \beta^2)} {\beta \, [\cosh (2\pi^2 \alpha^2 \beta^2) - \cos (2\pi \beta)]}. \nonumber
\end{align}
While the first function $F_p (d/D, D f_0 / v)$ depends on the precise form of $B_p (t)$ and only changes slowly with the transition frequency $f_0$, the second function $G (\tau / T, f_0 T)$ is more universal and features pronounced peaks at specific frequencies $f_0 = n / T$ that correspond to integer multiples of a fundamental frequency $1/T$ [see Fig.~\ref{fig-3}(b)]. Note that the width of each peak scales as $\Delta f_0 \sim n^2 \tau^2 / T^3$ and that the sharpest peak is thus the fundamental one corresponding to $n = 1$.

\emph{Discussion.}---For the proposed setup in Fig.~\ref{fig-1}, the transition frequency $f_0$ is constant, and the approximate periodicity $T$ is tuned by the voltage $V$ according to $T = \Phi / V$~\cite{Halperin-2010}. If we assume for simplicity that $\tau$ scales linearly with $T$ such that $\alpha = \tau / T$ is a constant, the total relaxation rate $\Gamma = \Gamma_p + \Gamma_q$ then takes the form [see Eq.~(\ref{eq-Gamma-4})]
\begin{equation}
\Gamma \approx \frac{\mu_B^2 \Phi^2 D^2 f_0} {\hbar^2 \lambda^4 v^2} \, F \left( \frac{d}{D}, \frac{D f_0} {v} \right) G \left( \alpha, \frac{\Phi f_0} {V} \right), \label{eq-Gamma-5}
\end{equation}
where $F (\gamma, \phi) = F_p (\gamma, \phi) + F_q (\gamma, \phi)$ becomes
\begin{align}
& F_{\textrm{NV}} (\gamma, \phi) = \frac{2}{3} \left[ F_x (\gamma, \phi) + F_y (\gamma, \phi) + F_z (\gamma, \phi) \right], \nonumber \\
& F_{\textrm{hBN}} (\gamma, \phi) = F_x (\gamma, \phi) + F_y (\gamma, \phi) \label{eq-F}
\end{align}
for NV centers and hBN vacancies that correspond to intrinsic field directions $\vec{u} = \{1,1,1\} / \sqrt{3}$ and $\vec{u} = \{0,0,1\}$, respectively. Note that $F_x (\gamma, \phi) = 0$ because of $B_x (t) = 0$.

The relaxation rate $\Gamma$ in Eq.~(\ref{eq-Gamma-5}) depends on the voltage $V$ through both the vortex speed $v$ and the dimensionless function $G (\alpha, \Phi f_0 / V)$. While the former dependence on $V$ is expected to be monotonic, the function $G (\alpha, \Phi f_0 / V)$ exhibits a sequence of characteristic peaks at voltages $V_n = \Phi f_0 / n$ (where $n$ is an integer), as shown in Fig.~\ref{fig-4}. Even though these peaks broaden and eventually disappear as $\alpha = \tau / T$ increases and the periodicity of the vortex motion becomes less perfect, the fundamental peak at voltage $V_1 = \Phi f_0$ remains clearly observable even for $\alpha = 0.2$.

The key difference between standard and half-quantum vortices is in the value of the magnetic flux $\Phi$ and, hence, the fundamental voltage $V_1$. In the case of standard vortices, $\Phi$ is simply the flux quantum $\Phi_0 \approx 2 \times 10^{-15}$ Tm$^2$ and, assuming $f_0 \approx 3$ GHz~\cite{Rondin-2014,Casola-2018,Gottscholl-2021,Liu-2022}, the fundamental peak corresponds to voltage $V_1 = \Phi_0 f_0 \approx$ $6$ $\mu$V. For half-quantum vortices carrying a smaller flux $\Phi = \Phi_0 / 2$, however, the fundamental voltage is reduced to $V_1 = \Phi_0 f_0 / 2 \approx$ $3$ $\mu$V.

From Eq.~(\ref{eq-Gamma-5}), the absolute value of the relaxation rate $\Gamma$ is on the order of $\mu_B^2 \Phi^2 D^2 f_0 / (\hbar^2 \lambda^4 v^2)$. In other words, the characteristic peaks described above are easier to observe in thicker superconducting films with shorter penetration depths $\lambda$ and slower vortex-crossing speeds $v$. Nevertheless, even for a relatively long penetration depth $\lambda \sim 1$ $\mu$m found in standard spin-triplet superconductor candidates~\cite{Aoki-2001,Bae-2021,Ishihara-2023}, reasonable parameter values $D \sim 100$ nm and $v \sim 10^4$ m/s~\cite{Bezuglyj-2022} translate into a relaxation time $T_1 = 1/\Gamma \lesssim 1$ ms that is resolvable with spin-qubit relaxometry~\cite{Degen-2017,Ariyaratne-2018,Wu-2025}. We remark that the spin-qubit distance $d$ must be kept relatively large, $d \gtrsim 100$ nm, so that vortices can be picked up even if they slightly deviate from their straight path~\cite{Footnote}. With $d/D \sim 1$ and $D f_0 / v \sim 0.01$, the function $F (d/D, D f_0 / v)$ is then of order unity, while the function $G (\alpha, \Phi f_0 / V)$ may reduce the relaxation time by another order of magnitude (see Fig.~\ref{fig-4}).

\emph{Summary.}---We proposed spin-qubit relaxometry as a feasible approach for measuring the magnetic fluxes carried by superconducting vortices that cross a narrow strip at a finite bias current. By matching the vortex-crossing frequency with the transition frequency of the spin qubit, the magnetic flux $\Phi$ of a single vortex is obtained as the ratio of the corresponding voltage and the known transition frequency. The extraction of a half-integer-quantized value, $\Phi = \Phi_0 / 2$, provides convincing evidence of both spin-triplet superconductivity and half-quantum vortices supporting Majorana zero modes.

\emph{Acknowledgments.}---This research was sponsored by the U.~S.~Department of Energy, Office of Science, Basic Energy Sciences, Materials Sciences and Engineering Division.




\begin{references}

\bibitem{Ivanov-2001} D. A. Ivanov, Phys. Rev. Lett. \textbf{86}, 268 (2001).
\bibitem{Alicea-2012} J. Alicea, Rep. Prog. Phys. \textbf{75}, 076501 (2012).
\bibitem{Kitaev-2001} A. Kitaev, arXiv:cond-mat/0010440.
\bibitem{Kitaev-2003} A. Y. Kitaev, Ann. Phys. \textbf{303}, 2 (2003).
\bibitem{Nayak-2008} C. Nayak, S. H. Simon, A. Stern, M. Freedman, and S. Das Sarma, Rev. Mod. Phys. \textbf{80}, 1083 (2008).
\bibitem{Ran-2019} S. Ran, C. Eckberg, Q.-P. Ding, Y. Furukawa, T. Metz, S. R. Saha, I.-L. Liu, M. Zic, H. Kim, J. Paglione, and N. P. Butch, Science \textbf{365}, 684 (2019).
\bibitem{Aoki-2022} D. Aoki, J.-P. Brison, J. Flouquet, K. Ishida, G. Knebel, Y. Tokunaga, and Y. Yanase, J. Phys. Condens. Matter \textbf{34}, 243002 (2022).
\bibitem{Fisher-1989} R. A. Fisher, S. Kim, B. F. Woodfield, and N. E. Phillips, L. Taillefer, K. Hasselbach, and J. Flouquet, A. L. Giorgi, and J. L. Smith, Phys. Rev. Lett. \textbf{62}, 1411 (1989).
\bibitem{Tou-1998} H. Tou, Y. Kitaoka, K. Ishida, K. Asayama, N. Kimura, Y. Onuki, E. Yamamoto, Y. Haga, and K. Maezawa, Phys. Rev. Lett. \textbf{80}, 3129 (1998).
\bibitem{Aoki-2001} D. Aoki, A. Huxley, E. Ressouche, D. Braithwaite, J. Flouquet, J.-P. Brison, E. Lhotel, and C. Paulsen, Nature \textbf{413}, 613 (2001).
\bibitem{Levy-2005} F. L\'evy, I. Sheikin, B. Grenier, and A. D. Huxley, Science \textbf{309}, 1343 (2005).
\bibitem{Mackenzie-2017} A. P. Mackenzie, T. Scaffidi, C. W. Hicks, and Y. Maeno, npj Quantum Materials \textbf{2}, 40 (2017).
\bibitem{Rondin-2014} L. Rondin, J.-P. Tetienne, T. Hingant, J.-F. Roch, P. Maletinsky, and V. Jacques, Rep. Prog. Phys. \textbf{77}, 056503 (2014).
\bibitem{Casola-2018} F. Casola, T. van der Sar, and A. Yacoby, Nat. Rev. Mater. \textbf{3}, 17088 (2018).
\bibitem{Gottscholl-2021} A. Gottscholl, M. Diez, V. Soltamov, C. Kasper, D. Krau{\ss}e, A. Sperlich, M. Kianinia, C. Bradac, I. Aharonovich, and V. Dyakonov, Nat. Commun. \textbf{12}, 4480 (2021).
\bibitem{Liu-2022} W. Liu, N.-J. Guo, S. Yu, Y. Meng, Z.-P. Li, Y.-Z. Yang, Z.-A. Wang, X.-D. Zeng, L.-K. Xie, and Q. Li, Mater. Quantum. Technol. \textbf{2}, 032002 (2022).
\bibitem{Degen-2017} C. L. Degen, F. Reinhard, and P. Cappellaro, Rev. Mod. Phys. \textbf{89}, 035002 (2017).
\bibitem{Agarwal-2017} K. Agarwal, R. Schmidt, B. I. Halperin, V. Oganesyan, G. Zar\'and, M. D. Lukin, and E. Demler, Phys. Rev. B \textbf{95}, 155107 (2017).
\bibitem{Rodriguez-Nieva-2018} J. F. Rodriguez-Nieva, K. Agarwal, T. Giamarchi, B. I. Halperin, M. D. Lukin, and E. Demler, Phys. Rev. B \textbf{98}, 195433 (2018).
\bibitem{Chatterjee-2019} S. Chatterjee, J. F. Rodriguez-Nieva, and E. Demler, Phys. Rev. B \textbf{99}, 104425 (2019).
\bibitem{Chatterjee-2022} S. Chatterjee, P. E. Dolgirev, I. Esterlis, A. A. Zibrov, M. D. Lukin, N. Y. Yao, and E. Demler, Phys. Rev. Research \textbf{4}, L012001 (2022).
\bibitem{Machado-2023} F. Machado, E. A. Demler, N. Y. Yao, and S. Chatterjee, Phys. Rev. Lett. \textbf{131}, 070801 (2023).
\bibitem{Potts-2025} M. Potts and S. Zhang, Nano Lett. \textbf{25}, 17677 (2025).
\bibitem{Finco-2021} A. Finco, A. Haykal, R. Tanos, F. Fabre, S. Chouaieb, W. Akhtar, I. Robert-Philip, W. Legrand, F. Ajejas, K. Bouzehouane, N. Reyren, T. Devolder, J.-P. Adam, J.-V. Kim, V. Cros, and V. Jacques, Nat. Commun. \textbf{12}, 767 (2021).
\bibitem{Kumar-2024} J. Kumar, D. Yudilevich, A. Smooha, I. Zohar, A. K. Pariari, R. St{\"o}hr, A. Denisenko, M. H{\"u}cker, and A. Finkler, Nano Lett. \textbf{24}, 4793 (2024).
\bibitem{Melendez-2025} A. L. Melendez, S. Das, F. Ayala Rodriguez, I.-H. Kao, W. Liu, A. J. Williams, B. Lv, J. Goldberger, S. Chatterjee, S. Singh, and P. C. Hammel, Sci. Adv. \textbf{11}, eadu9381 (2025).
\bibitem{Wu-2025} Y.-C. Wu, G. B. Hal\'asz, J. T. Damron, Z. Gai, H. Zhao, Y. Sun, K. A. Dahmen, C. Sohn, E. W. Carlson, C. Hua, S. Lin, J. Song, H. N. Lee, and B. J. Lawrie, Nano Lett. \textbf{25}, 1473 (2025).
\bibitem{Li-2025} Y. Li, Z. Ding, C. Wang, H. Sun, Z. Chen, P. Wang, Y. Wang, M. Gong, H. Zeng, F. Shi, and J. Du, Nat. Commun. \textbf{16}, 8585 (2025).
\bibitem{Ariyaratne-2018} A. Ariyaratne, D. Bluvstein, B. A. Myers, and A. C. Bleszynski Jayich, Nat. Commun. \textbf{9}, 2406 (2018).
\bibitem{Monge-2023} R. Monge, T. Delord, N. V. Proscia, Z. Shotan, H. Jayakumar, J. Henshaw, P. R. Zangara, A. Lozovoi, D. Pagliero, P. D. Esquinazi, T. An, I. Sodemann, V. M. Menon, and C. A. Meriles, Nano Lett. \textbf{23}, 422 (2023).
\bibitem{Jayaram-2025} S. Jayaram, M. Lenger, D. Zhao, L. Pupim, T. Taniguchi, K. Watanabe, R. Peng, M. Scheffler, R. St\"ohr, M. S. Scheurer, J. Smet, and J. Wrachtrup, Phys. Rev. Lett. \textbf{135}, 126001 (2025).
\bibitem{Liu-2025} Z. Liu, R. Gong, J. Kim, O. K. Diessel, Q. Xu, Z. Rehfuss, X. Du, G. He, A. Singh, Y. S. Eo, E. A. Henriksen, G. D. Gu, N. Y. Yao, F. Machado, S. Ran, S. Chatterjee, and C. Zu, arXiv:2502.04439.
\bibitem{Halperin-2010} B. I. Halperin, G. Refael, and E. Demler, Int. J. Mod. Phys. B \textbf{24}, 4039 (2010).
\bibitem{Bezuglyj-2022} A. I. Bezuglyj, V. A. Shklovskij, B. Budinsk\'a, B. Aichner, V. M. Bevz, M. Yu. Mikhailov, D. Yu. Vodolazov, W. Lang, and O. V. Dobrovolskiy, Phys. Rev. B \textbf{105}, 214507 (2022).
\bibitem{Bae-2021} S. Bae, H. Kim, Y. S. Eo, S. Ran, I. Liu, W. T. Fuhrman, J. Paglione, N. P. Butch, and S. M. Anlage, Nat. Commun. \textbf{12}, 2644 (2021).
\bibitem{Ishihara-2023} K. Ishihara, M. Roppongi, M. Kobayashi, K. Imamura, Y. Mizukami, H. Sakai, P. Opletal, Y. Tokiwa, Y. Haga, K. Hashimoto, and T. Shibauchi, Nat. Commun. \textbf{14}, 2966 (2023).
\bibitem{Footnote} We anticipate the extent of such spatial deviations to be determined by the size of the smaller constriction at which the vortices are nucleated. This size can be kept below $100$ nm.

\end{references}
\end{document}